\documentstyle[12pt,epsfig]{article}

\catcode`\@=11
\long\def\@makefntext#1{
\protect\noindent \hbox to 3.2pt {\hskip-.9pt  
$^{{\ninerm\@thefnmark}}$\hfil}#1\hfill}                

\def\@makefnmark{\hbox to 0pt{$^{\@thefnmark}$\hss}}  
        
\def\ps@myheadings{\let\@mkboth\@gobbletwo
\def\@oddhead{\hbox{}
\rightmark\hfil\ninerm\thepage}   
\def\@oddfoot{}\def\@evenhead{\ninerm\thepage\hfil
\leftmark\hbox{}}\def\@evenfoot{}
\def\sectionmark##1{}\def\subsectionmark##1{}}

\renewcommand{\thefootnote}{\fnsymbol{footnote}}

\newcounter{sectionc}\newcounter{subsectionc}\newcounter{subsubsectionc}
\renewcommand{\section}[1] {\vspace*{0.6cm}\addtocounter{sectionc}{1} 
\setcounter{subsectionc}{0}\setcounter{subsubsectionc}{0}\noindent 
        {\normalsize\bf\thesectionc. #1}\par\vspace*{0.4cm}}
\renewcommand{\subsection}[1] {\vspace*{0.6cm}\addtocounter{subsectionc}{1} 
        \setcounter{subsubsectionc}{0}\noindent 
        {\normalsize\it\thesectionc.\thesubsectionc. #1}\par\vspace*{0.4cm}}
\renewcommand{\subsubsection}[1]
{\vspace*{0.6cm}\addtocounter{subsubsectionc}{1}
        \noindent {\normalsize\rm\thesectionc.\thesubsectionc.\thesubsubsectionc. 
        #1}\par\vspace*{0.4cm}}

\newcounter{appendixc}
\newcounter{subappendixc}[appendixc]
\newcounter{subsubappendixc}[subappendixc]

\renewcommand{\appendix}[1] {\vspace*{0.6cm}
        \refstepcounter{appendixc}
        \setcounter{figure}{0}
        \setcounter{table}{0}
        \setcounter{equation}{0}
        \renewcommand{\thefigure}{\Alph{appendixc}.\arabic{figure}}
        \renewcommand{\thetable}{\Alph{appendixc}.\arabic{table}}
        \renewcommand{\theappendixc}{\Alph{appendixc}}
        \renewcommand{\theequation}{\Alph{appendixc}.\arabic{equation}}
        \noindent{\bf Appendix \theappendixc #1}\par\vspace*{0.4cm}}

\def\sta{\tilde{t}_1}
\def\stb{\tilde{t}_2}

\def\hb{\mbox{$h^0$~}}
\def\Ab{\mbox{$A^0$~}}
\def\MH{\mbox{$M_H$~}}
\def\Mh{\mbox{$M_h$~}}
\def\MA{\mbox{$M_A$~}}

\def\aeff{\mbox{$\alpha_{eff}$~}}

\def\ra{\rightarrow}

\def\sqrts{\mbox{$\surd{s}$~}}

\def\abstracts#1{{
        \centering{\begin{minipage}{12.2truecm}\footnotesize\baselineskip=12pt\noindent
        \centerline{\footnotesize ABSTRACT}\vspace*{0.3cm}
        \parindent=0pt #1
        \end{minipage}}\par}} 

\renewenvironment{thebibliography}[1]
        {\begin{list}{\arabic{enumi}.}
        {\usecounter{enumi}\setlength{\parsep}{0pt}
\setlength{\leftmargin 1.25cm}{\rightmargin 0pt}
         \setlength{\itemsep}{0pt} \settowidth
        {\labelwidth}{#1.}\sloppy}}{\end{list}}

\newcounter{itemlistc}
\newcounter{romanlistc}
\newcounter{alphlistc}
\newcounter{arabiclistc}

\newcommand{\fcaption}[1]{
        \refstepcounter{figure}
        \setbox\@tempboxa = \hbox{\footnotesize Fig.~\thefigure. #1}
        \ifdim \wd\@tempboxa > 6in
           {\begin{center}
        \parbox{6in}{\footnotesize\baselineskip=12pt Fig.~\thefigure. #1}
            \end{center}}
        \else
             {\begin{center}
             {\footnotesize Fig.~\thefigure. #1}
              \end{center}}
        \fi}

\newcommand{\tcaption}[1]{
        \refstepcounter{table}
        \setbox\@tempboxa = \hbox{\footnotesize Table~\thetable. #1}
        \ifdim \wd\@tempboxa > 6in
           {\begin{center}
        \parbox{6in}{\footnotesize\baselineskip=12pt Table~\thetable. #1}
            \end{center}}
        \else
             {\begin{center}
             {\footnotesize Table~\thetable. #1}
              \end{center}}
        \fi}

\def\@citex[#1]#2{\if@filesw\immediate\write\@auxout
        {\string\citation{#2}}\fi
\def\@citea{}\@cite{\@for\@citeb:=#2\do
        {\@citea\def\@citea{,}\@ifundefined
        {b@\@citeb}{{\bf ?}\@warning
        {Citation `\@citeb' on page \thepage \space undefined}}
        {\csname b@\@citeb\endcsname}}}{#1}}

\newif\if@cghi
\def\cite{\@cghitrue\@ifnextchar [{\@tempswatrue
        \@citex}{\@tempswafalse\@citex[]}}
\def\citelow{\@cghifalse\@ifnextchar [{\@tempswatrue
        \@citex}{\@tempswafalse\@citex[]}}
\def\@cite#1#2{{$\null^{#1}$\if@tempswa\typeout
        {IJCGA warning: optional citation argument 
        ignored: `#2'} \fi}}

 1
 1
 1

\font\ninerm=cmr9

\begin{document}

\centerline{\normalsize\bf RADIATIVE CORRECTIONS IN THE MSSM HIGGS SECTOR}

\vspace*{0.6cm}
\centerline{\footnotesize WOLFGANG HOLLIK}
\baselineskip=13pt
\centerline{\footnotesize\it Institut f\"ur Theoretische Physik, Universit\"at
Karlsruhe}
\baselineskip=12pt
\centerline{\footnotesize\it D-76128 Karlsruhe, Germany}
\centerline{\footnotesize E-mail: hollik@itpaxp3.physik.uni-karlsruhe.de}

\vspace*{0.9cm}
\abstracts{1-loop diagrammatic calculations of cross sections and
            decay widths of neutral Higgs bosons in the minimal 
            supersymmetric standard model 
            are reviewed and compared with 
             compact expressions in the effective potential approximation.}
 
\normalsize\baselineskip=15pt
\setcounter{footnote}{0}
\renewcommand{\thefootnote}{\alph{footnote}}

\section{Introduction}

In order to experimentally detect possible signals from the 
Higgs sector of the minimal supersymmetric standard model (MSSM),
detailed studies for the decay and production processes of Higgs bosons are 
required.
As discovered several years ago~\cite{GT,EPA,RGE},
radiative corrections in the MSSM Higgs sector are large and have to be 
taken into account for phenomenological studies. Three main approaches
have been developed to calculate the 1-loop 
radiative corrections to the MSSM Higgs
boson masses, production and decay rates:
\begin{itemize}
\item[(i)] The Effective Potential Approach (EPA)~\cite{EPA}.
\item[(ii)] The method of    
            Renormalization Group Equations (RGE)~\cite{RGE}. 
\item[(iii)] The diagrammatic calculation in the on-shell renormalization 
scheme (Feynman Diagram Calculation, 
FDC)~\cite{BRI,MYNPB,dabelstein,decayew,dh,DHR}:
 The masses are calculated from the pole positions of 
the Higgs propagators, and the cross sections are obtained from the full set
of 1-loop diagrams contributing to the amplitudes.
\end{itemize}
\noindent
The method (iii) is technically involved, but it is the most accurate one
at the 1-loop level and can be used as a reference frame
for simpler approximations.
The searches for Higgs bosons at LEP~\cite{lepexp} and studies 
for the future searches at higher energies~\cite{highen} conventionally make 
use of the very compact formulation in the
effective potential approximation. 

This talk gives an overview on the neutral MSSM Higgs 
sector at the 1-loop level. The results for 
the cross sections of neutral Higgs production processes 
in $e^+e^-$ collisions and for the neutral Higgs decay widths
are discussed in a complete diagrammatic calculation
 and compared, where possible, 
with the corresponding 
ones of the compact EPA approximation.

\newpage
 
\section{One-loop calculations}

The tree level potential for the neutral MSSM Higgs bosons can be
written as follows:
\begin{eqnarray}
V = m_1^2 H_1^2 +  m_2^2 H_2^2 
+ \epsilon_{ij}(m_{12}^2 H_1^i H_2^j + H.c.)
+ {g^2 + {g'}^2\over 8}(H_1^2-H_2^2)^2 + {g^2\over 4}(H_1 H_2)^2\, .
\label{eq:pot0}
\end{eqnarray}
Diagonalization of the mass matrices for the CP-even and the CP-odd
scalars, following from the potential~(\ref{eq:pot0}), leads to three
physical particles: two CP-even Higgs bosons
$H^0$,\hb and one CP-odd
Higgs boson \Ab, and defines their tree-level masses $m_h, m_H,m_A$  
and the mixing angles $\alpha, \beta$.
For a sytematic 1-loop caclulation, the free parameters of the Higgs
potential  $m_1^2,\, m_2^2,\, m_{12}^2,\, g,\, g'$
and the two vacua $v_1,\, v_2$ are replaced by renormalized parameters
plus counter terms. This transforms the potential $V$ into 
$V +\delta V$, where V, expressed in the renormalized parameters,
is formally identical to (\ref{eq:pot0}), and 
$ \delta V $
is the counter term potential. The counter terms are fixed by seven
renormalization conditions. In the on-shell scheme they can be chosen as 
follows:  
\begin{itemize}
\item the on-shell conditions for $M_{W,Z}$ and the electric charge $e$ as
in the minimal standard model. 
\item the on-shell condition for the $A^0$ boson with the pole mass $M_A$.
\item the tadpole conditions for vanishing renormalized tadpoles:
$$ T_H + \delta t_H = 0, \quad 
    T_h + \delta t_h = 0 
$$
where $T_{H,h}$ are the sum of the 1-loop tadpole diagrams for $H^0$ and 
$h^0$, and  $\delta t_{H,h}$ are the tadpole counter terms following from (1).
These conditions ensure that $v_1,\, v_2$ are the minima of
the  potential at the 1-loop level. 
\item the renormalization of $\tan\beta $ in such a way 
that the relation $\tan\beta= v_2/v_1$ is valid for the 1-loop 
Higgs minima.
\end{itemize}
By this set of conditions, the input for the
MSSM Higgs sector is fixed by $M_A$ and $\tan\beta$, together
with the standard gauge sector input $M_{W,Z}$ and $e$. 
The last condition on $\tan\beta$ can only be imposed in connection with an
appropriate field renormalization of the two Higgs doublet fields.
Together with the gauge field renormalization one has
four extra renormalization
constants which can be fixed as in the standard model gauge sector 
\cite{bhs}, extended by two more conditions for the Higgs sector. 
The latter two
have been treated in two slightly different ways 
\cite{MYNPB,dabelstein}
in the literature;
physical results, however, differ only
marginally by unobservably small terms.
The corresponding 1-loop physical Higgs boson masses 
$M_h^2, M_H^2$ are
obtained as the pole positions  of the dressed scalar propagators. 

\bigskip 
In the EPA, the tree level potential is improved by adding the
1-loop terms~.
The 1-loop potential $V^{(1)}$ is rediagonalized yielding the 1-loop
corrected physical masses $M_H, M_h$ and the effective mixing angle
\aeff.
These improved masses and \aeff are used in the Born formulae for
production and decay rates of Higgs bosons. In the approximation keeping 
only  the dominating terms $\sim m_t^4$, the expressions read \cite{EPA}:

\begin{eqnarray}
 M^2_{H,h} & = & \frac{M_A^2 + M_Z^2 + \epsilon_t + \sigma_t}{2}
 \pm \, \left[ \
 \frac{ (M_A^2 + M_Z^2)^2 + ( \epsilon_t - \sigma_t )^2}{4}
 - M_A^2 M_Z^2 \cos^2 2\beta \right. \nonumber \\ & & \left. + \
 \frac{(\epsilon_t - \sigma_t) \cos 2\beta}{2}  (M_A^2 - M_Z^2)
 - \lambda_t \sin 2\beta (M_A^2 + M_Z^2) + \lambda_t^2 \
 \right]^{1/2} 
\label{glmapp2} \nonumber \\
\end{eqnarray}
with
\begin{eqnarray}
 \epsilon_t = \frac{N_C G_F m_{t}^4 }{\sqrt{2} \pi^2 \sin^2 \beta} \
  & & \hspace{-0.5cm}
 \left[  \log \ ( \frac{m_{\tilde{t}_1} m_{\tilde{t}_2} }{m_{t}^2} )
 + \frac{A_t ( A_t + \mu \cot \beta)}{ m_{\tilde{t}_1}^2 -
 m_{\tilde{t}_2}^2 } \log \frac{ m_{\tilde{t}_1}^2 }{ m_{\tilde{t}_2}^2 }
 \right. \nonumber \\ & &  \hspace*{-0.5cm}    \left. +
 \frac{A_t^2 ( A_t + \mu \cot \beta)^2 }{ ( m_{\tilde{t}_1}^2 -
 m_{\tilde{t}_2}^2 )^2} \left( 1 - \frac{m_{\tilde{t}_1}^2 + m_{\tilde{t}_2}^2}
 {m_{\tilde{t}_1}^2 - m_{\tilde{t}_2}^2} \log \frac{m_{\tilde{t}_1}}
 {m_{\tilde{t}_2}} \right) \ \right]   \nonumber \\
 \lambda_t = \frac{N_C G_F m_{t}^4 }{\sqrt{2} \pi^2 \sin^2 \beta} \
  & & \hspace{-0.5cm}
 \left[ \frac{ \mu ( A_t + \mu \cot \beta)}{ m_{\tilde{t}_1}^2 -
 m_{\tilde{t}_2}^2 } \log \frac{ m_{\tilde{t}_1}^2 }{ m_{\tilde{t}_2}^2 }
 \right. \nonumber \\ & &  \hspace*{-0.5cm}    \left. +
 \frac{2 \mu A_t ( A_t + \mu \cot \beta)^2 }{ ( m_{\tilde{t}_1}^2 -
 m_{\tilde{t}_2}^2 )^2} \left( 1 - \frac{m_{\tilde{t}_1}^2 + m_{\tilde{t}_2}^2}
 {m_{\tilde{t}_1}^2 - m_{\tilde{t}_2}^2} \log \frac{m_{\tilde{t}_1}}
 {m_{\tilde{t}_2}} \right) \ \right]   \nonumber \\
 \sigma_t = \frac{N_C G_F m_{t}^4 }{\sqrt{2} \pi^2 \sin^2 \beta} \
  & & \hspace{-0.5cm}
 \frac{ \mu^2 ( A_t + \mu \cot \beta)^2 }{ ( m_{\tilde{t}_1}^2 -
 m_{\tilde{t}_2}^2 )^2} \left[ 1 - \frac{m_{\tilde{t}_1}^2 + m_{\tilde{t}_2}^2}
 {m_{\tilde{t}_1}^2 - m_{\tilde{t}_2}^2} \log \frac{m_{\tilde{t}_1}}
 {m_{\tilde{t}_2}} \right]  \ .
\label{leadom2}
\end{eqnarray}
The approximate effective mixing angle $\alpha_{eff}$ is determined by
\begin{equation}
 \tan \alpha_{eff} = \frac{ - (M_A^2 + M_Z^2) \ \sin\beta  \cos\beta\ 
                           + \lambda_t }{
 M_Z^2 \cos^2\beta + M_A^2 \sin^2 \beta \ + \ \sigma_t - \ M_h^2 } \ .
\label{aleff}
\end{equation}
These formulae contain the masses $m_{\tilde{t}_{1,2}}$ of the top squarks,
the Higgs mixing parameter $\mu$ of the superpotential, and the non-diagonal 
entry  $A_t$ in the stop mass matrix. This matrix is diagonal for
$A_t+\mu \cot\beta =0$. In this special case we
have $\sigma_t = \lambda_t = 0$, and only the $\epsilon_t$ term contributes.

\medskip 
Recently the leading 2-loop corrections
to the CP-even MSSM Higgs boson masses have been investigated, based on
the EPA and RGE methods~\cite{twoloop}. 
The main conclusion is that 2-loop corrections are also significant and 
tend to compensate partially the effects of the 1-loop corrections.

\section{Production cross sections for 
  $e^+e^-\rightarrow Z h^0(H^0), \, A^0 h^0 (H^0)$   }

In this section the results for $Z^0H^0(Z^0h^0)$ and 
$A^0H^0(A^0h^0)$ production
 are shown as derived   
from the complete 1-loop FDC, and  the
quality of the corresponding EPA results is discussed.
The formulae for the cross sections obtained in the FDC differ from the
Born expressions 
not only by the corrections to the masses and to the angle $\alpha$,
but also by new form factors and momentum dependent effects
(see~\cite{MYNPB,dh} for analytic expressions).

\begin{figure}[htbp]
\centerline{\epsfig{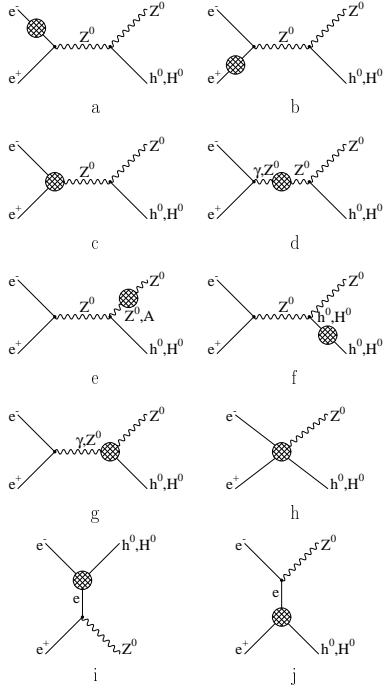}}
\vspace*{-0.5cm}
\caption{\small       
 Classes of diagrams contributing to the 
  $e^+e^-\ra Z^0h^0(H^0)$ process in the FDC approach. 
\label{fig:feyn}}
\end{figure}

For the calculations of the cross sections we need the full set of 
2-, 3- and 4-point functions. In Fig.~\ref{fig:feyn} the 
diagrams contributing
to  $e^+e^-\ra Z^0h^0(H^0)$ are collected. The  diagrams 
for $e^+e^-\ra A^0h^0(H^0)$ can be obtained by
changing $Z^0$ into $A^0$ on the external line and 
skipping the diagrams i), j).

In the figures of this section the set of parameters (in GeV): 
$M_A=200,\, M_2=1000,\, \mu=500,\, M_{sl}=300,\, M_{sq}=1000,\,
 A_t=A_b=1000$
is used as an example.
$\mu$ is the parameter describing the Higgs 
doublet mixing in the MSSM superpotential. $M_2$ denotes the SU(2) gaugino 
mass parameter. For the U(1) gaugino mass we use the value 
$M_1=\frac{5}{3} \tan^2\theta_W M_2$, suggested by GUT constraints.  
$M_{sq},M_{sl},A_t$ and $A_b$ are the parameters entering the sfermion mass 
matrices.
For simplicity we assume a common value $M_{sq}$ for all generations 
of squarks, and a common $M_{sl}$ for sleptons.

\begin{figure}[htbp]
\begin{tabular}{p{0.48\linewidth}p{0.48\linewidth}}             
\begin{center}  
\mbox{     
\psfig{file=fig.4a,width=\linewidth,bbllx=0pt,bblly=25pt,bburx=520pt,bbury=520pt}}
\end{center}  & 
\begin{center}  
\mbox{          
\psfig{file=fig.4b,width=\linewidth,bbllx=0pt,bblly=25pt,bburx=520pt,bbury=520pt}}
\end{center} \\ 
\end{tabular}   
\caption{
\small
\baselineskip=12pt
Comparison of the cross sections \mbox{$\sigma(e^+e^-\ra Z^0h^0,
A^0h^0)$}
obtained in the EPA and FDC.  \sqrts = 500 GeV.
\label{fig:crl500}
}               
\end{figure}
In the conventional  $M_A, \tan\beta$ parametrization,
Fig.~\ref{fig:crl500} shows the production cross sections 
$\sigma(e^+e^-\ra Z^0h^0, A^0h^0)$ 
at $\sqrts = 500$ GeV.
For the choosen set of parameters the numerical differences can reach 30\% at
$\sqrts=500$ GeV.  Note, however, that in the region of large
cross sections the EPA accuracy is better (20\% at 500 GeV). 
The situation of $H^0$ production is very similar.
The differences between EPA and FDC  
become more important with increasing energies, 
exceeding 40\% at 1 TeV. Also the
effect of the additional form factors in the FDC grows, which
modify the angular dependence 
of the cross section compared to the 
effective Born approximation.
More detailed discussions
can be found in ref.~\cite{DHR,highen}.

A more physical parametrization of the cross sections is given in 
terms of the two Higgs boson masses \MA and \Mh (or \MH), instead of the 
formal quantity $\tan\beta$. 
This parametrization is more involved 
in the calculations, but it has the advantage of physically well defined
input quantities avoiding possible confusions from different renormalization 
schemes. Varying e.g.~\MH (\MA and other input quantities fixed)
we obtain $\tan\beta$ and $\sigma_{ZH}$, $\sigma_{AH}$ as functions of
\MH. 
Significant differences can occur for the cross sections, as displayed 
in Fig.~\ref{fig:tbcr} where the predictions of EPA and FDC for the
$\sigma_{ZH}$ and $\sigma_{AH}$ are plotted as functions of \MH. 
The typical size of the differences between the methods is 
10-20\% for $\sqrts=500$ GeV, but they may became quite large (60\%) for
the process $\sigma(e^+e^- \ra Z^0 H^0)$. In other cases, they are 
of the order 10--20\%.

\begin{figure}[htbp]
\begin{tabular}{p{0.48\linewidth}p{0.48\linewidth}}             
\begin{center}  
\mbox{     
\psfig{file=fig.8a,width=\linewidth,bbllx=0pt,bblly=25pt,bburx=520pt,bbury=520pt}}
\end{center}  & 
\begin{center}  
\mbox{          
\psfig{file=fig.8b,width=\linewidth,bbllx=0pt,bblly=25pt,bburx=520pt,bbury=520pt}}
\end{center} \\ 
\end{tabular}   
\caption{
\small
\baselineskip=12pt
Comparison of the cross sections \mbox{$\sigma(e^+e^- \ra Z^0 H^0, A^0
H^0)$} versus
 \MH in the EPA and FDC. \sqrts = 500 GeV.
\label{fig:tbcr}
}               
\end{figure}

The variation of the
SUSY parameters: sfermion and gaugino masses, $\mu$
parameter and sfermion mixing parameters,
does not
have a large effect on the size of the differences between the EPA and FDC.
Hence, the figures represent typical examples.

\smallskip
Summarizing this section, 
comparisons between the FDC and EPA predictions
have shown that at $\sqrts=500$ GeV
the EPA has an accuracy of typically 10-20\% in the 
parameter regions where the cross sections are large.
The differences become larger with increasing energy, where also modifications
of the Born-like angular distributions are more visible. 
The use of the physical input variables $M_A,$ $M_h$ or $M_A,$ $M_H$ 
avoids ambiguities
from the definition of $\tan\beta$ in higher order, but the observed
differences remain of the same size.
For a better accuracy, the full FDC would be 
required. 

\smallskip
So far the leading 2-loop terms have not been incorporated.
They would improve the 1-loop FDC results in the same way as the 
approximations and thus do not influence the remaining differences which can 
only be obtained by an explicit diagrammatic calculation.

\bigskip 
Loop-induced pair production 
$e^+e^- \rightarrow h^0 h^0, H^0 H^0, h^0 H^0, A^0 A^0$
of neutral Higgs bosons
have also been studied recently \cite{loopinduced},
as well as associated Higgs--photon production
$e^+e^- \rightarrow \gamma h^0 (H^0,A^0)$ \cite{hgamma}.
The general result is that the cross sections in the MSSM 
 are not enhanced by the extra non-standard particles in the loops,
$\sigma(MSSM) \leq \sigma(SM)$.
In the decoupling limit, for heavy SUSY particles, the standard model
results are recovered. For detailed studies of the virtual SUSY effects 
in the loop-induced Higgs-$\gamma Z$ and Higgs-$\gamma\gamma$
couplings see ref.~\cite{hgz}.

\newpage

\section{Decays of neutral Higgs bosons}

The decay widths (the branching ratios, respectively) as well as the
mass--width correlations are quantities which 
can help to differentiate between
Higgs bosons of different origin. Except for a small region of the parameter 
space, the light neutral Higgs of the MSSM decays predominantly into
$b$-quarks and $\tau$-leptons; the heavier ones $H^0$ and $A^0$ can 
have significant decay modes also into top quarks, scalar quarks, and
neutralinos/charginos. In a certain region of the parameter space,
also the decay $h^0 \rightarrow A^0 A^0$ is allowed.
Loop-mediated decay processes are the 
hadronic decay modes into gluons \cite{gluons} and gluinos
\cite{gluinos}.

\subsection{Fermionic decays}

For the important fermionic decays both electroweak
 and QCD corrections \cite{decayew,decayqcd,coarasa}
have been calculated. The standard QCD corrections \cite{decayqcd} for 
$\phi \rightarrow q \bar{q}$, 
$\phi = h^0, H^0, A^0$, 
 are large. The bulk can be  absorbed into the running quark mass by
 replacing the pole mass according to  
 $m_q \rightarrow  \overline{m}_q(M_{\phi})$.
The SUSY-QCD corrections arising from virtual gluinos and squarks 
\cite{decayew,coarasa} can also become remarkably large,
in particular for large
values of $\mu$ and $\tan\beta$ where they can reach up to 30\%.

\begin{figure}[htbp]
\vspace*{-2.cm}
\centerline{\psfig{figure=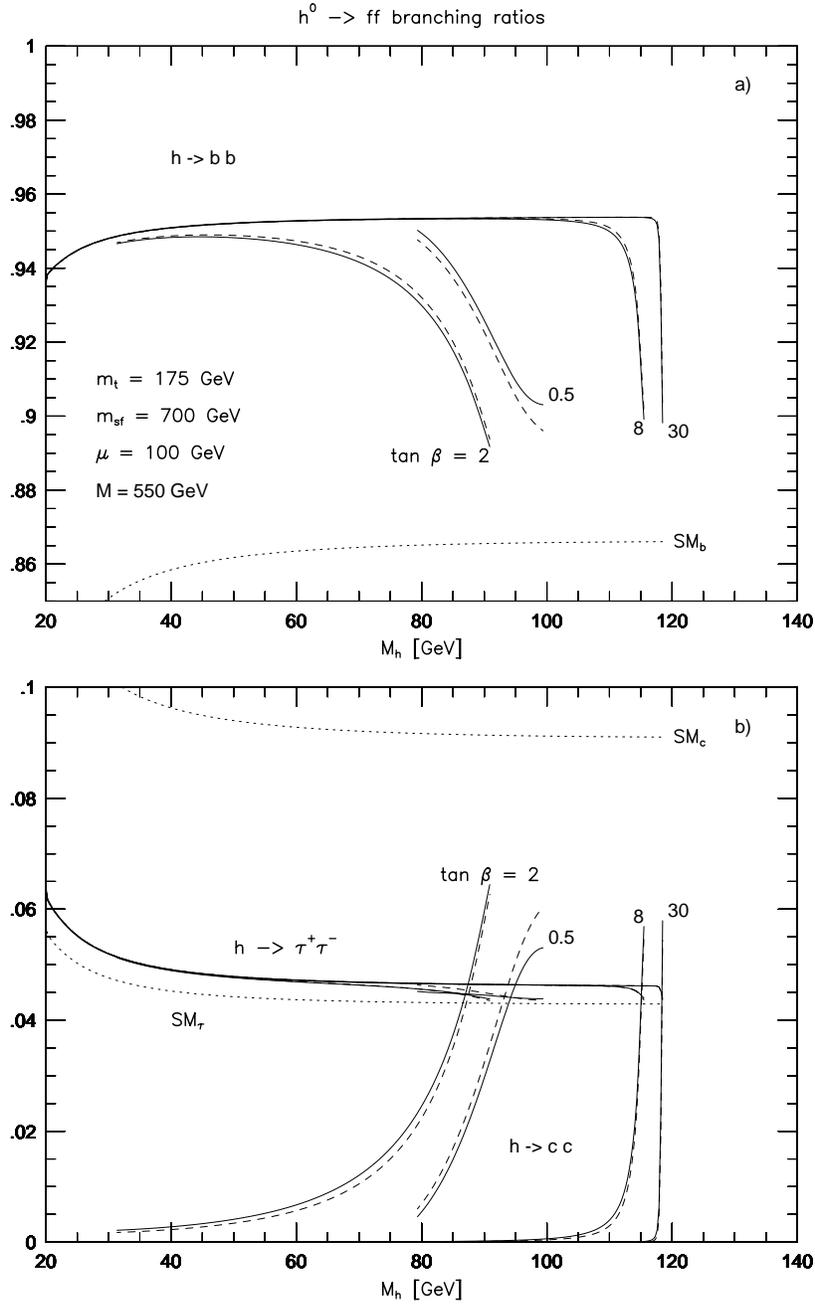,height=21.5cm}}
\vspace*{-1cm}
\caption {\small Fermionic branching ratios
Higgs decays into fermion pairs (from {\protect\cite{decayew}}).
EPA: dashed lines; complete 1-loop: full lines.
$M$ denotes the SU(2) gaugino mass.}
\label{fermbr}
\end{figure}

\smallskip
The set of 1-loop electroweak corrections
to $h^0\rightarrow f\bar{f}$  can be 
summarized in terms the following decay amplitude:
\begin{equation}
 A(h\rightarrow f\bar{f}) \ = \ \sqrt{Z_h} \left(
 \Gamma_h - \frac{\Sigma_{hH}(M_h^2)}{M_h^2-m_H^2+\Sigma_{HH}(M_h^2)} \
 \Gamma_H \right) 
\label{hff}
\end{equation}   
with the renormalized self-energies $\Sigma$ and 3-point vertex functions 
$\Gamma_{h,H}$.
The amplitude for $H^0\rightarrow f\bar{f}$ is obtained 
by interchanging $h\leftrightarrow H$. 
The wave function renormalization $Z_{h(H)}$ is the finite residue of the
$h^0\, (H^0)$ propagator. The amplitude for $A^0\rightarrow f\bar{f}$
is given by the renormalized vertex $\Gamma_A$ alone,
due to the renormalization condition $Z_A = 1$ \cite{dabelstein}.
For the absolute decay widths, in order to have the correct normalization
in terms of the Fermi constant $G_F$, the MSSM correction to the 
muon lifetime, i.e.~the quantity $\Delta r$, has to be taken into account
\cite{deltar}. It drops out in the branching ratios.
The inclusion of the mixing term in eq.~(\ref{hff})
corresponds essentially to the 
rediagonalization of the mass matrix in the EPA. 
In the EPA, one obtains the improved decay amplitude by using the 
EPA masses and the effective mixing angle \aeff, eq.~(\ref{aleff}), 
in the Born expression for the vertex $\Gamma_h$, with $Z_h=1$ and 
$\Sigma_{hH}=0$ in eq.~(\ref{hff}).
As shown in \cite{decayew}, the EPA is a very good approximation
of the full 1-loop result 
for the fermionic branching ratios,
with exception of extremely low values for
$\tan\beta$ (see Fig.~\ref{fermbr}).

\subsection{The decay $h^0 \rightarrow A^o A^0$}

A specific consequence of the large radiative corrections to the
$h^0$ mass is the possibility of having
$M_h > 2 M_A$, which makes the decay $h^0 \rightarrow A^o A^0$
kinematically allowed at 1-loop order for low values of $\tan\beta$
and $M_A$. In the allowed region it turns out to be the 
dominant decay mode, with branching ratios of 0.8--0.9. 
The decay width was obtained 
in the EPA (last reference of \cite{EPA}), 
by use of the RGE \cite{haa-rge}, and 
by a complete diagrammatic 1-loop calculation \cite{haa}.
A simple approximate formula for the decay amplitude, which
reproduces the full 1-loop decay width at an accuracy within typically
10\%, is given by \cite{haa}
\begin{equation}
A(h^0\rightarrow A^0 A^0) \ = \ - \ \frac{g}{2 c_W}\ M_Z \ \cos 2\beta 
                       \ \sin(\alpha_{eff}+\beta) \ + \ \Delta V 
\label{hAA}
\end{equation}
with \aeff from (\ref{aleff}) and 
\begin{eqnarray}
\Delta V & \simeq  & \frac{3 g^3}{16 \pi^2 M_W^3} \ 
               \frac{\cos\alpha_{eff}}{\sin\beta} \
                \cot^2\beta \ \left[ 
        m_t^4 \ \log \frac{m_t^2}{m_{\sta} m_{\stb}} \right.  \nonumber \\
  &  & \left.    \  -\ (A_t-\mu\tan\beta)^2 \ m_t^4 \
            \frac{\log m_{\sta} - \log m_{\stb} }{m_{\sta}^2 - m_{\stb}^2 }
   \ + \   m_t^2 \ \frac{M_h^2-2 M_A^2}{2} \  \right] 
\label{deltaV}
\end{eqnarray}
which arises from the vertex correction diagrams with virtual top and
stop quarks. It contributes significantly to the decay amplitude in
particular for $\tan\beta \simeq 1$, where the improved Born approximation
[the first term in eq.~(\ref{hAA})] is very small.
The second term in $\Delta V$ is also sizeable for intermediate
masses of the top squarks. For details see ref.~\cite{haa}.

\subsection{Higgs decays into squark pairs}

In a large part of the MSSM parameter space, the decays 
into squark pairs
$H^0, A^0 \rightarrow \tilde{q} \overline{\tilde{q}}$  
can be the dominant decay modes \cite{hss}.
QCD corrections from gluons and gluinos have been calculated \
recently \cite{hssqcd}. They are sizeable (up to 50\%) and should be
taken into account for phenomenological studies.


\vskip 0.4 cm 
{\normalsize \bf \noindent References}
\vskip 0.3 cm


\begin{thebibliography}{99}

\bibitem{GT} S. P. Li, M. Sher, {\sl Phys.~Lett.}~{\bf B140}
(1984), 339;
J. Gunion, A. Turski, {\sl Phys.~Rev.}~{\bf D39} (1989) 2701; 
{\bf D40} (1989) 2325, 2333;
M. Berger,  {\sl Phys.~Rev.}~{\bf D41} (1990) 225.

\bibitem{EPA} J. Ellis, G. Ridolfi,  F. Zwirner,
{\sl Phys.~Lett.}~{\bf 262B} (1991) 477;
R. Barbieri, M. Frigeni, {\sl Phys.~Lett.}~{\bf 258B} (1991) 395;
A. Brignole, J. Ellis, G. Ridolfi, F. Zwirner
{\sl Phys.~Lett.}~{\bf 271B} (1991) 123.

\bibitem{RGE}
      H.E. Haber, R. Hempfling, {\sl Phys.~Rev.~Lett.}~{\bf 66} (1991) 1815; 
       {\sl Phys.~Rev.}~{\bf D48} (1993) 4280;
       M. Carena, K. Sasaki,  C.E.M. Wagner,
      {\sl Nucl.~Phys.}~{\bf 381B} (1992)~66.
     P.H. Chankowski, S. Pokorski,  J. Rosiek, 
                         {\em Phys.~Lett.}~{\bf 281B} (1992) 100.

\bibitem{BRI} A. Brignole, {\sl Phys.~Lett.}~{\bf 281B} (1992) 284.

\bibitem{MYNPB} P. Chankowski, S. Pokorski, J. Rosiek 
{\sl Nucl.~Phys.}~{\bf B423} (1994) 437; {\bf B423} (1994) 497.
 
\bibitem{dabelstein} 
 A. Dabelstein, W. Hollik, in: {\sl $e^+e^-$  Collisions at 500 GeV},
 DESY 92-123C, ed.~P. Zerwas;
 A. Dabelstein, {\sl Z. Phys.}~{\bf C67} (1995) 495.

\bibitem{decayew} A. Dabelstein, {\sl Nucl.~Phys.}~{\bf B456} (1995) 25.

\bibitem{dh}
V. Driesen, W. Hollik, {\sl Z. Phys.}~{\bf C68} (1995) 485.

\bibitem{DHR} V. Driesen, W. Hollik, J. Rosiek,
        {\sl  Z. Phys.}~{\bf C71} (1996) 259.

\bibitem{lepexp} P. M\"attig, 
 {\sl 28th International Conference on High
Energies}, Warsaw 1996 (plenary talk);
{\sl Higgs Physics}, M. Carena, P. Zerwas et al., in:
{\sl Physics at LEP2}, CERN 96-01, CERN 1996, 
eds.~G. Altarelli, T. Sj\"ostrand, F. Zwirner.

\bibitem{highen}  
{\sl Higgs Particles}, 
in: {\sl $e^+e^-$  Collisions at 500 GeV}, 
DESY 92-123A,B,C,D, ed.~P. Zerwas. 

\bibitem{bhs} M. B\"ohm, W. Hollik, H. Spiesberger,
              {\sl Fortschr.~Phys.}~{\bf 34} (1986) 687;
              W. Hollik,
              {\sl Fortschr.~Phys.}~{\bf 38} (1990) 165.
  





\bibitem{twoloop} R. Hempfling, A. Hoang, {\sl Phys.~Lett.}~{\bf 331B} (1994) 
99;
M. Carena, J.R. Espinosa, M. Quiros, C.E.M. Wagner, 
{\sl Phys.~Lett.}~{\bf 355B} (1995) 209; 
M. Carena, M. Quiros, C.E.M. Wagner,{\sl Nucl.~Phys.}~{\bf B461} (1996) 407.


\bibitem{loopinduced} A. Djouadi, V. Driesen, C. J\"unger,
                      {\sl Phys.~Rev.}~{\bf D54} (1996) 759.

\bibitem{hgamma} A. Djouadi, V. Driesen, W. Hollik, J. Rosiek,
                hep-ph/9609420, to appear in {\sl Nucl.~Phys.}~B.

\bibitem{hgz} A. Djouadi, V. Driesen, W. Hollik, J.I. Illana,
              hep-ph/9612362;
              A. Djouadi, V. Driesen, W. Hollik, A. Kraft,
              hep-ph/9701342.

\bibitem{gluons} A. Djouadi, M. Spira, P. Zerwas,
                 {\sl Phys.~Lett.}~{\bf 264B} (1991) 440;
                 M. Spira et al., {\sl Nucl.~Phys.}~{\bf B453} (1995) 17. 

\bibitem{gluinos} A. Djouadi, M. Drees, 
                  {\sl Phys.~Rev.}~{\bf D51} (1995) 4997.


\bibitem{decayqcd} E. Braaten, J.P. Leveille, 
                  {\sl Phys.~Rev.}~{\bf D22} (1980) 715; D. Bardin et al.,
                  {\sl Yad.~Fiz.}~{\bf 53} (1991) 240;
               M. Drees, K. Hikasa, {\sl Phys. Lett.}~{\bf 276B} (1990) 455. 

\bibitem{coarasa} J.A. Coarasa, R.A. Jimenez, J. Sol\`a, 
                  {\sl Phys.~Lett.}~{\bf 389B} (1996) 312.
                   
\bibitem{deltar} P. Chankowski et al., {\sl Nucl.~Phys.}~{\bf B417} (1994) 101;
                 D. Garcia, J. Sol\`a, {\sl Mod.~Phys.~Lett.}~{\bf A9}
                                       (1994) 211. 

\bibitem{haa-rge} H.E. Haber, R. Hempfling, Y. Nir,
                  {\sl Phys.~Rev.}~{\bf D46} (1992) 3015.

\bibitem{haa} S. Heinemeyer, W. Hollik, {\sl Nucl.~Phys.}~{\bf B474}
                                        (1996) 32. 

\bibitem{hss} A. Djouadi et al., hep-ph/9605339; A. Bartl et al., 
                 hep-ph/9607388.

\bibitem{hssqcd} A. Bartl et al., hep-ph/9701398;
                 A. Arhrib, A. Djouadi, W. Hollik, C. J\"unger,
                 hep-ph/9702426. 

\end{thebibliography}
\end{document}